\PassOptionsToPackage{fleqn}{amsmath}
\documentclass[twocolumn,preprintnumbers,amsmath,amssymb,floatfix]{revtex4}
\usepackage{graphicx}
\usepackage{dcolumn}
\usepackage{bm}
\usepackage{amssymb}
\usepackage{amsmath}

\begin{document}
\title{Synthesis of Perovskite SrIrO3 Thin Films by Sputtering Technique}

\author{Z.Z. Li$^{1}$, O. Schneegans$^{2}$, L. Fruchter$^{1}$}%
\affiliation{$^1$Laboratoire de Physique des Solides, C.N.R.S., Universit\'{e} Paris-Sud, 91405 Orsay, France}
\affiliation{$^2$Laboratoire G\'{e}nie Electrique et Electronique de Paris, CNRS UMR 8507, UPMC and Paris-Saclay Universities, Centralesup\'{e}lec, 91192 Gif Sur Yvette, France}

\begin{abstract}{We report on the synthesis of perovskite SrIrO3 thin films using sputtering technique. Single phase (110) oriented SrIrO$_3$ thin films were epitaxially grown on SrTiO3 (001) substrate. Using off-axis XRD $\theta-2\theta$ scans, we demonstrate that these films exhibit (110) out-of-plane orientation with (001) and (1-10) lying in-plane. The sputtering grown thin films have a smooth, homogeneous surface, and excellent coherent interface with the substrate.}
\end{abstract}
\maketitle

\section*{\label{intro}Introduction}

In the last decade, a new epitaxial perovskite SrIrO$_3$ (hereafter SIO)  has been grown as thin films and drawn great attention due to its unique attractive properties. SIO is a thermally and chemically stable oxide material with good electrical conductivity. Its perovskite structure is compatible with many superconducting, ferroelectric materials, and perovskite substrates\cite{Becher14,Junquera03,Yi03}, which makes SIO a good candidate as a new class of electrode material for microelectronic devices used under harsh environmental conditions. Polycrystalline bulk samples only can be grown under high pressure, and epitaxy was found the only way to obtain the single crystal material\cite{Longo71}. SIO thin films have been grown using molecular beam epitaxial\cite{Nie15,Liu16c}, pulsed laser deposition (PLD)\cite{Moon08,Jang10,Wu13,Liu13,Biswas14, Liu16b,Gruenewald14,Biswas15}, and metalorganic chemical vapor deposition\cite{Kim06}. To the best of our knowledge, it has not been reported that SIO thin films can be grown using the sputtering technique, which is recognized as a high-reproducibility and large-area thin film fabrication technique. The conditions to elaborate SIO thin film by this technique is thus of great importance for its potential industrial applications. In this paper, we report for the first time the synthesis of high quality SIO thin films by this technique. A structural characterization using x-ray diffraction (XRD) technique reveals their crystal structure information and growth orientation. The transport properties of these films are reported elsewhere\cite{Fruchter16}.

\section{\label{Experimental}Experimental}

The single crystal-like orthorhombic SrIrO$_3$ thin films have been prepared by means of on-axis geometry, single target, radio frequency (RF) magnetron sputtering on heated, single crystal (001) SrTiO$_3$ substrates  (hereafter: STO). We used the chemically stable compound\cite{Powell93} Sr$_4$IrO$_6$ as the sputtering target, to compensate for the observed reduction of the thin films Sr content, by a factor $\approx$ 4. The target was prepared using a conventional solid state method. Briefly, stoichiometric IrO$_2$ (99.99 \% in purity) and dried SrCO$_3$ (99.99\% in purity) were mixed and pre-reacted at 700 $^\circ$C in an alumina crucible for 24 hours to decompose the carbonate. The resulting materials were reground, pelletized, and fired successively for three days at 1100 $^\circ$C, and for another four days at 1200 $^\circ$C. The diameter of the target was  50 mm and the power applied on the target was 30 W during the thin film growth. Before being mounted on the substrate holder with a silver paste, the substrates were cleaned in an ultrasonic bath with acetone and alcohol for 15 minutes, sequentially. Then the substrates were exposed under plasma environment in a plasma cleaner at room temperature for 3 minutes. During the SIO thin film deposition, the temperature of the substrates was kept at 610 $^\circ$C, monitored by an infrared pyrometer. The sputtering gas was a mixture of Ar and O$_2$ with ratio of 1 to 2 and the total pressure was fixed at 210 mTorr, with a 40 mm target-to-substrate distance. To remove any possible contamination on the surface of the target during charge and discharge of the samples, five minutes pre-sputtering was carried out prior to the deposition of the thin films. The thin films were found stable when stored in ambient atmosphere, even for films as thin as a few nm (\textit{a contrario}, Ref.~\cite{Groenendijk16} reports the necessity to encapsulate PLD samples for stability). We could grow films as thick as 100 nm, without any detectable parasitic phase, in particular monoclinic SrIrO$_3$\cite{Zhang13}. However, while  60 nm films were found single crystal, off-axis XRD revealed that the 100 nm film was twinned in the substrate plane.

\begin{figure}
\resizebox{0.95\columnwidth}{!}{%
  \includegraphics{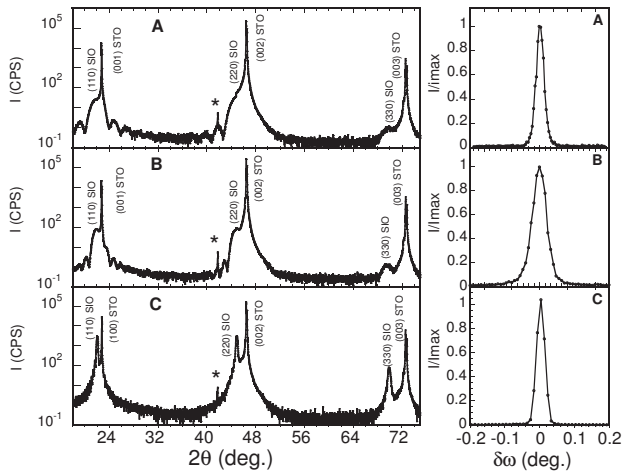}
  }
\caption{a: XRD $\theta-2\theta$ scans of three SrIrO$_3$ films with different thickness (film A, B, and C), deposited on (001) STO substrate. ($^*$) is from the $\kappa_\beta$ (002)STO reflection. b: rocking curves around the (220) reflection of SIO thin films for A, B and C films.}\label{fig1}
\end{figure}

To assess the quality and crystallographic orientation of newly grown SIO thin films, XRD measurements were performed using a PANAlytical X'Pert thin film diffractometer. The XRD diffraction patterns were obtained using Cu $K_\alpha$ ($\lambda$= 1.5418 \AA) radiation operated at 40 kV and 20 mA. The fixed divergence slit of 1/32 deg. and the Soller slit with a divergence of 0.04 radians were employed. A parallel plate collimator and a flat graphite monochromator crystal (a slit of 0.1 mm in-between) were adopted during XRD analysis. We carried out $\theta$-2$\theta$ scan and rocking curve for phase identification and crystal quality assessment. The thin film thickness is determined from the period of the Kiessig fringes around the SIO diffraction peaks in the XRD $\theta$-2$\theta$ scan, and by a classic X-ray reflectivity (XRR) technique. The high-pressure form of SrIrO$_3$ has a crystal structure of perovskite SrRuO$_3$-type, with orthorhombic p$_{bnm}$ symmetry\cite{Longo71,Zhao08, Puggioni16}. Its lattice constants are \textit{a} = 5.56 \AA, \textit{b} = 5.59 \AA, and \textit{c} = 7.88 \AA. To determine the growth orientation of the SIO layer on cubic (001) STO substrate, we used the off-axis XRD technique. Finally, to assess the epitaxial growth quality of SIO thin films, the reciprocal space maps (RSM) measurements were used. 

\section{Results and discussion}

Figure ~\ref{fig1} shows the XRD $\theta$-2$\theta$ scans of SIO thin films, with three different thicknesses (named A, B, and C) grown on (001) STO. Film thicknesses vary by changing only the deposition time, while keeping all other film growth conditions fixed. It is noticed that only (00l) diffraction peaks from cubic STO substrate and (hh0) diffraction peaks from  the SIO orthorhombic structure are detected, indicating the formation of a pure phase of SrIrO3 layer. The reason we index the SIO layer with out-of-plane of [110], but not [001], will be discussed in detail later. The out-of-plane (110) spacing value 4.038 \AA  is determined for the film C. This value is larger than the one given for bulk samples \cite{Longo71,Zhao08}, and agrees with what was reported in Reference \onlinecite{Biswas14}, for films grown using the PLD technique. This originates from the mismatch value of -1.5\% between SIO and (001) STO \cite{Biswas14}, as the in-plane compression strain along [1-10] and [001] directions leads to tensile strain along the out-of-plane [110] direction.

\begin{figure}
\resizebox{0.95\columnwidth}{!}{%
  \includegraphics{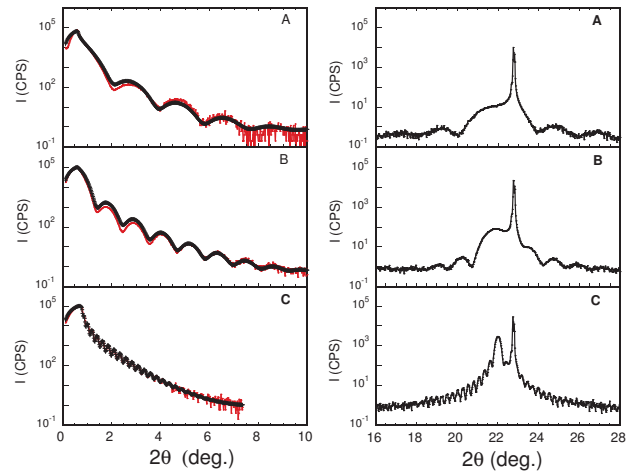}
  }
\caption{a : XRR data (red line) and simulation (black crosses) of the films A, B and C. b :  enlarged view of the (110) peak in figure \ref{fig1}a .}\label{fig2}
\end{figure}

\begin{figure}
\resizebox{0.6\columnwidth}{!}{%
  \includegraphics{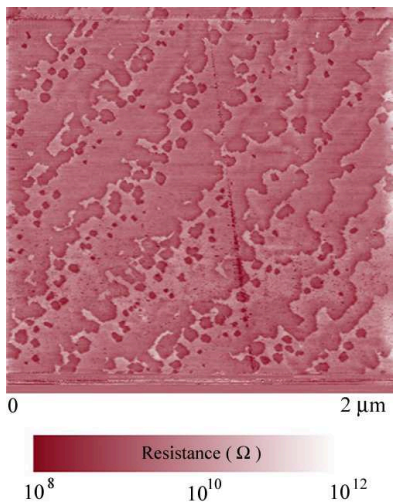}
  }
\caption{Contact resistance AFM image, for a 63 nm thick film.}\label{fig0}
\end{figure}

The thicknesses of these three SIO films are determined by the XRR technique. The spectra of XRR measurement illustrated in the figure ~\ref{fig2} were fitted with a PANAlytical commercial software. The measured thicknesses of these three films obtained in this way  are 4.6 (Film A), 7.7 (Film B), and 34 nm (Film C). A second thickness measurement  was also obtained from the Kiessig fringes. Figure ~\ref{fig2}b shows the enlarged region scan in Figure ~\ref{fig1}. The regularity of thickness fringes can be seen in the $\theta-2\theta$ scan around SIO diffraction peaks for all three films. Using the expression $ t = \lambda/(2x\Delta\omega_f \cos(\theta)$ (where $\Delta\omega_f$ is the difference between adjacent thickness fringe angles and $\theta$ is Bragg reflection angle), the measured thicknesses of these three SIO films are found 4.5, 7.6, and 34 nm, respectively, in close agreement with the values obtained from XRR measurements. It is known that these Kiessig oscillations result from the coherent scattering from a finite number of lattice planes, thus carrying crystalline thin film thickness information. Hence, it indicates the homogeneous nature of the surface of these films\cite{Biswas15}. The atomic flatness of grown thin films is also confirmed by AFM measurements. Fig.~\ref{fig0} shows the contact resistance of an AFM scan, for a thick (63 nm) film. Steps are indicative of a 2D growth (step flow), and it was checked that the step height is $\approx$ 0.4 nm, i.e. the pseudo-cubic cell parameter. The rocking curves in Figure ~\ref{fig1}b are taken from the (220) reflection. The full width at half maximum (FWHM) is found 0.07, 0.06 and 0.035 deg., for film A,B and C, respectively. In the same configuration, the rocking curve for the (002) reflection of STO is about 0.03 deg., which is close to the value of the intrinsic peak broadening of the instrument. The very small values of the rocking curves FWHM for all films indicates an excellent out-of-plane alignment to the substrates.

The growth orientation was determined, using the off-axis XRD $\theta$-2$\theta$ scans. Indeed, the standard $\theta-2\theta$ scans do not allow to discriminate the (hh0) peaks from the close (00l=2h) ones. This method is widely used to determine two in-plane axis directions of the (110) RScO$_3$ single crystal substrate (R: rare earth). Figure 3 shows four XRD $\theta$-2$\theta$ scans with rotation angles  $\phi$ = 0, 90, 180, and 270 deg., corresponding to the scans along the STO $\left\{202\right\}$ planes, with sample tilted at angle $\Psi$ = 45 deg. From the spectra, we can see that the four STO (202) diffraction peaks are perfectly superposed on each other, which indicates excellent accuracy of our XRD instrument. For $\phi$ = 0 deg. and 180 deg., the two diffraction peaks from SIO thin film are superposed, corresponding to (224) and (22-4), if we assume it has [110] out-of-plane orientation. For $\phi$ = 90 deg. and 270 deg., the two peaks are not superposed to the (224) and (22-4) peaks, and can be indexed as the (400) and (040) reflections of the SIO film, in agreement also with [110] out-of-plane orientation of the film. Therefore, by using these simple $\theta$-2$\theta$ scans with the 45 deg. sample tilting method, we can obtain orientation information about the grown SIO thin films and conclude: i/ the out-of-plane orientation of the SIO films is  the [110] direction; ii/ in-plane (400) and (040) can be easily distinguished with interplanar distance d$_{(400)}$ = 5.60 \AA  and d$_{(040)}$ = 5.64 \AA. We further determine d$_{(004)}$ = 7.84 \AA \cite{Note1}. It was found that the substrate vicinality systematically orients the in-plane orientation of the films, with the [001]direction lying nearest to the vicinal steps one. A similar growth orientation was reported for SIO films on (001) STO, elaborated with PLD, as evidenced by  transmission electronic microscopy direct observations\cite{Zhang13}, and for SIO grown on GdScO$_3$\cite{Liu16b}, from a complete structure determination. It has been reported that SrRuO$_3$ (SRO) epitaxial layers are grown on (001) STO with out-of-plane [110] direction and (001) and (1-10) aligned in-plane\cite{Gan99,Vailionis08}. The authors proposed a growth mechanism to account for the [110] out-of-plane orientation of the SRO film on STO. As SRO and SIO crystals have the same crystallographic structure with similar lattice parameters, it is not surprising that SrIrO$_3$ epitaxial layers are grown in the [110] out-of-plane direction also.

\begin{figure}
\resizebox{0.95\columnwidth}{!}{%
  \includegraphics{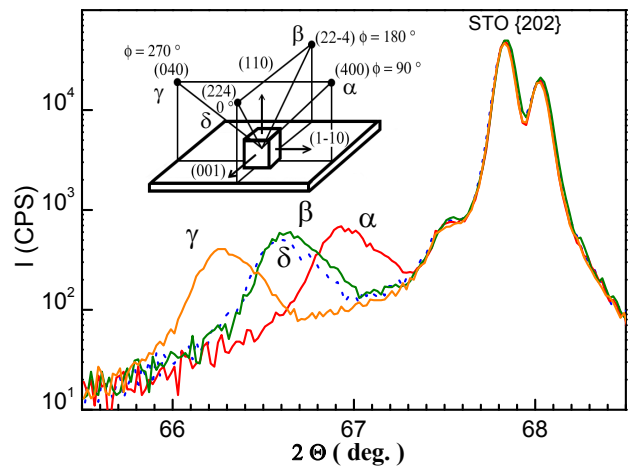}
  }
\caption{Four off-axis $\theta-2\theta$ scans of the SIO Film C along STO {202}, with in-plane rotation angle of $\phi$ = 0, 90, 180 and 270 deg. and  tilting angle of $\Psi$ = 45 deg. The sketch displays the four superstructure peaks, labeled respectively $\alpha$, $\beta$, $\gamma$, $\delta$.}\label{fig3}
\end{figure}

\begin{figure}
\resizebox{0.95\columnwidth}{!}{%
  \includegraphics{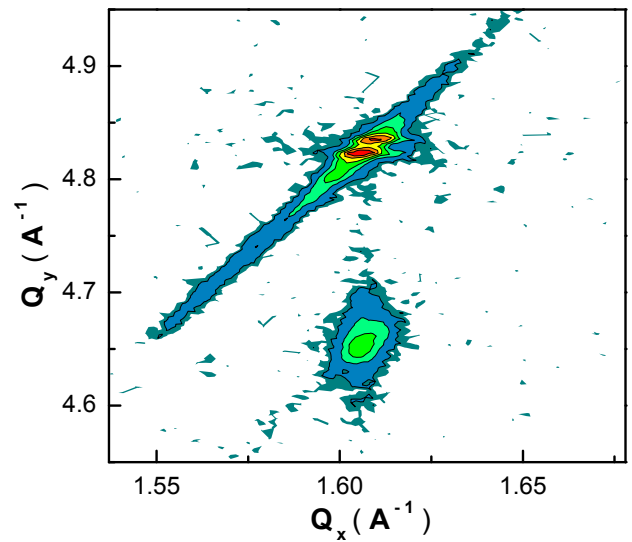}
  }
\caption{Reciprocal space mapping (RSM) of pseudo-cubic SIO (103) and STO (103) reflections.}\label{fig4}
\end{figure}

Figure \ref{fig4} shows the RSM data  for the (332) diffraction ray ((103) in the pseudo-cubic notation often used in the literature), for the thicker 34 nm film.  The reciprocal space map (RSM) measurements are carried out around (103) reflection for either SIO layer or STO substrate. It can be seen that the Q$_x$ value for the SIO film and the one for the STO substrate are identical, indicating that the SIO film is coherently constrained on the substrates.

In summary, we demonstrated that high quality SIO thin films can be synthesized using RF sputtering technique, with a single non-stoichiometric target. The single crystal-like SrIrO$_3$ thin films with atomically flat surface have an excellent out-of-plane alignment, and are fully coherently constrained in-plane on STO substrates. Using off-axis XRD $\theta$-2$\theta$ scans, we demonstrated that the films exhibit (110) out-of-plane orientation with (001) and (1-10) lying on the substrate surface.

\section*{}

We acknowledge support from the Agence Nationale de la Recherche grant SOCRATE, and valuable help from G. Collin in the handling of crystallographic data.


\end{document}